\documentclass[twocolumn,secnumarabic,amssymb, nobibnotes, aps, prd,
  superscriptaddress, preprintnumbers]{revtex4-1} 
\usepackage{epsfig}
\usepackage{longtable}

\newcommand{\gA}{g_{\rm{A}}}
\newcommand{\rmO}{{\textrm{O}}}
\newcommand{\rme}{{\rm{e}}}
\newcommand{\Nf}{N_{\rm{f}}}
\newcommand{\fm}{{\rm{fm}}}
\newcommand{\MeV}{{\rm{MeV}}}
\newcommand{\csw}{c_{\rm{sw}}}

\begin{document}

\preprint{\texttt{MKPH-T-12-11}}
\preprint{\texttt{HIM-2012-02}}
\preprint{\texttt{CERN-PH-TH-2012-106}}
\title{The nucleon axial charge in lattice QCD with controlled errors} 

\author{S. Capitani}
\affiliation{Institut f\"ur Kernphysik, University of Mainz, Becher
  Weg 45, D-55099 Mainz, Germany}
\affiliation{Helmholtz Institute Mainz, University of Mainz, D-55099
  Mainz, Germany} 
\author{M. Della Morte}
\affiliation{Institut f\"ur Kernphysik, University of Mainz, Becher
  Weg 45, D-55099 Mainz, Germany}
\affiliation{Helmholtz Institute Mainz, University of Mainz, D-55099
  Mainz, Germany}
\author{G. von Hippel}
\affiliation{Institut f\"ur Kernphysik, University of Mainz, Becher
  Weg 45, D-55099 Mainz, Germany}
\author{B. J\"ager}
\affiliation{Institut f\"ur Kernphysik, University of Mainz, Becher
  Weg 45, D-55099 Mainz, Germany}
\affiliation{Helmholtz Institute Mainz, University of Mainz, D-55099
  Mainz, Germany} 
\author{A. J\"uttner}
\affiliation{CERN, Physics Department, Theory Division, CH-1211
  Geneva 23, Switzerland}
\author{B. Knippschild}
\affiliation{Institut f\"ur Kernphysik, University of Mainz, Becher
  Weg 45, D-55099 Mainz, Germany}
\author{H.B. Meyer}
\affiliation{Institut f\"ur Kernphysik, University of Mainz, Becher
  Weg 45, D-55099 Mainz, Germany}
\author{H. Wittig}
\email[Email:]{wittig@kph.uni-mainz.de}
\affiliation{Institut f\"ur Kernphysik, University of Mainz, Becher
  Weg 45, D-55099 Mainz, Germany}
\affiliation{Helmholtz Institute Mainz, University of Mainz, D-55099
  Mainz, Germany} 

\date{September 2012}%


\begin{abstract}
We report on our calculation of the nucleon axial charge $\gA$ in QCD
with two flavours of dynamical quarks. A detailed investigation of
systematic errors is performed, with a particular focus on
contributions from excited states to three-point correlation
functions. The use of summed operator insertions allows for a much
better control over such contamination. After performing a chiral
extrapolation to the physical pion mass, we find
$\gA=1.223\pm0.063\,({\rm stat}){}^{+0.035}_{-0.060}\,({\rm syst})$,
in good agreement with the experimental value.
\end{abstract}

\maketitle

\section{Introduction}

Lattice simulations of Quantum Chromodynamics (QCD) have, by
now, reached a stage which allows for first-principles determinations
of many hadronic properties, with overall uncertainties at the percent
level\,\cite{Colangelo:2010et}. While systematic errors for quantities
such as quark masses, meson decay constants and form factors appear
very well controlled, the situation regarding properties of the
nucleon is less satisfactory. For instance, lattice calculations have
so far failed in reproducing the well-known experimental findings on
nucleon structure (see \cite{Renner_lat09,alexandrou_lat10} for recent
reviews). A prominent example is the axial charge, $\gA$, of the
nucleon. Lattice results for this quantity lie typically $10-15$\,\%
below the experimental value \cite{nuclFF:LHPC02_nf2,
nuclFF:RBC_lat04_nf2, nuclFF:LHPC05_nf2p1, nuclFF:QCDSF06_nf2,
nuclFF:RBC08_nf2p1, nuclFF:RBC08_nf2, nuclFF:RBC09_nf2p1,
nuclFF:BGR_lat09, nuclFF:QCDSF_lat09, nuclFF:QCDSF_lat10,
nuclff:QCDSF_lat10gA, nuclFF:LHPC10_nf2p1, nuclff:ETMC10_nf2}. What is
even more worrying is the absence of any tendency in the lattice data
which would indicate that the gap is narrowing as the pion mass is
decreased --- in fact, the opposite trend is often observed. The most
likely explanation is that systematic effects are not fully
controlled. What is lacking, therefore, is a benchmark calculation of
a quantity which describes basic structural properties of the nucleon,
and for this purpose the axial charge is an ideal candidate: (1) it is
derived from a matrix element of a simple fermionic bilinear which
contains no derivatives, (2) the initial and final states can both be
considered at rest, and (3) its definition as an iso-vector quantity
implies that contributions from quark-disconnected diagrams are
absent.

In this paper, we report on our results for $\gA$ addressing in detail
all sources of systematic errors, such as lattice artefacts,
finite-volume effects, and chiral extrapolations. We specifically
focus on the problem of a systematic bias arising from excited state
contributions in the relevant correlation functions. To this end, we
apply the method of summed operator insertions, which helps to control
any such contamination.

\section{Simulation details}

Our simulations are performed with $\Nf=2$ flavours
of $\rmO(a)$ improved Wilson fermions and the Wilson plaquette action. We
stress that excited state contamination is an important issue for lattice
simulations with any number of dynamical quarks. Hence, the question whether
estimates for $\gA$ may be biased can be adequately addressed in two-flavour
QCD. In particular, there is ample evidence\,\cite{Colangelo:2010et} that
there are no discernible differences between QCD with $\Nf=2$ and $\Nf=2+1$
flavours at the few-percent level. Therefore, the observed gap between
previous lattice estimates of the axial charge and its experimental value is
by far too large to be explained by the presence or absence of a dynamical
strange quark.

We use the non-perturbative determination of the improvement coefficient
$\csw$ from ref.\,\cite{impr:csw_nf2}. Table\,\ref{tab_params} contains a
compilation of lattice sizes and other simulation parameters, including the
pion and nucleon masses in lattice units. All listed ensembles were generated
as part of the CLS initiative, employing the deflation-accelerated DD-HMC
algorithm \cite{DDHMC,DDHMC-defl}. Quark propagators were computed using
Gaussian-smeared source vectors \cite{smear:Gaussian89} supplemented by
HYP-smeared links\,\cite{smear:HYP01}. The smearing parameters were tuned to
maximize plateau lengths for effective masses for a variety of channels. On
each ensemble we collected between 150 and 250 highly decorrelated
configurations. Up to eight sources, equally spaced in the temporal direction,
were used to reduce statistical fluctuations in correlation functions.  In
this way, we performed between 280 and 1700 individual measurements on our
ensembles. The lattice spacings were determined using the mass of the $\Omega$
baryon as described in\,\cite{GvH_lat11}. As we are in the process of
supplementing the set of our ensembles, estimates of the lattice spacing will
be updated in the future. Correlation functions were computed using the same
smeared nucleon interpolating operators at the source and sink. For
three-point functions we employed the improved axial current which is related
to its renormalized counterpart via \cite{impr:pap1}
\begin{equation}
  (A_\mu^{\rm R}) = Z_{\rm A}(1+b_{\rm A}am_q)(A_\mu
   +ac_{\rm A}\partial_\mu P),
\label{eq:za}
\end{equation}
where $A_\mu$ and $P$ denote the local axial current and pseudoscalar density,
respectively, and $m_q$ is the bare subtracted quark mass. Since $\gA$ was
determined from the $3^{\rm{rd}}$ component of the axial current alone, the
contribution proportional to $\partial_\mu P$ vanishes, as the axial charge is
defined at zero momentum transfer. Non-perturbative values for the
renormalization factor $Z_{\rm A}$ were taken from
ref.\,\cite{impr:za_nf2upd}, while the improvement coefficient $b_{\rm{A}}$
was estimated in tadpole-improved perturbation theory\,\cite{impr:pap5}. Since
the contribution from the improvement term is at the sub-percent level in the
range of quark masses considered, the systematic effect arising from the
unknown non-perturbative value for $b_{\rm{A}}$ will be negligible.

\begin{table}
\begin{ruledtabular}
\begin{tabular}{c c c c c c c c c}
Run & $L/a$ & $\beta$ & $\kappa$ & $am_\pi$ & $am_{\rm N}$ &
$m_{\pi}L$ & $N_{\rm cfg}$ & $N_{\rm src}$ \\
\hline
A2 & 32 & 5.2 & 0.13565 & 0.2424( 4) & 0.592(4) & 7.73 & 144 & 4 \\ 
A3 & 32 & 5.2 & 0.13580 & 0.1893( 5) & 0.531(4) & 6.06 & 265 & 4 \\
A4 & 32 & 5.2 & 0.13590 & 0.1454( 7) & 0.481(6) & 4.65 & 199 & 4 \\
A5 & 32 & 5.2 & 0.13594 & 0.1249(14) & 0.469(8) & 4.00 & 212 & 8 \\
\hline
E3 & 32 & 5.3 & 0.13605 & 0.2071( 6) & 0.510(3) & 6.63 & 139 & 2 \\
E4 & 32 & 5.3 & 0.13610 & 0.1934( 5) & 0.497(3) & 6.19 & 162 & 8 \\
E5 & 32 & 5.3 & 0.13625 & 0.1439( 6) & 0.420(3) & 4.60 & 168 & 8 \\
F6 & 48 & 5.3 & 0.13635 & 0.1036( 5) & 0.382(5) & 4.97 & 199 & 4 \\
F7 & 48 & 5.3 & 0.13538 & 0.0886( 4) & 0.334(8) & 4.25 & 250 & 4 \\
\hline
N4 & 48 & 5.5 & 0.13650 & 0.1358( 3) & 0.351(2) & 6.52 & 150 & 4 \\
N5 & 48 & 5.5 & 0.13660 & 0.1090( 3) & 0.320(3) & 5.23 & 150 & 4 \\
\end{tabular}
\end{ruledtabular}
\caption{Simulation parameters, pion and nucleon masses for all ensembles used
  in this paper. The temporal extent of each lattice is twice the
  spatial length, $T=2L$. $N_{\rm cfg}$ is the number of
  configurations per ensemble, while $N_{\rm src}$ denotes the number
  of different sources.}
\label{tab_params}
\end{table}

\section{Excited state contamination}

We denote the Euclidean time separation between the
nucleon source and sink by $t_{\rm s}$, while $t$ with $t\leq t_{\rm
s}$ marks the interval between the source and the axial current. If,
as in our case, the same smeared interpolating operators are applied
at the source and sink, the axial charge can be determined from a
simple ratio,
\begin{equation}
R(t,t_{\rm s}) := \frac{C_3^{\rm A}(t,t_{\rm s})}{C_2(t_{\rm s})},
\end{equation}
where $C_3^{\rm A}(t,t_{\rm s})$ denotes the three-point correlation
function of the local, bare axial current at vanishing momentum
transfer. For large values of $t$ and $t_{\rm s}$ the ratio
$R(t,t_{\rm s})$ yields directly the bare axial charge, i.e. 
\begin{equation}
  R(t,t_{\rm s}) 
  \stackrel{t,(t_{\rm s}-t)\gg0}{\longrightarrow}\; \gA^{\rm{bare}}
  +\rmO(\rme^{-\Delta t}) +\rmO(\rme^{-\Delta(t_{\rm s}-t)}),
\label{eq:Rratio}
\end{equation}
where $\Delta$ denotes the gap between the masses of the nucleon and its first
excitation. The axial charge is usually extracted by fitting $R(t,t_{\rm s})$
to a constant. Due to the exponentially increasing noise-to-signal ratio in
correlation functions of the nucleon, typical values of $t_{\rm s}$ are of the
order of 1\,\fm. To guarantee a reliable determination of $\gA$, excited state
contributions in eq.\,(\ref{eq:Rratio}) must already be sufficiently
suppressed for $t, (t_{\rm s}-t)\lesssim 0.5\,\fm$.

\begin{figure}
\begin{center}
\includegraphics[width=8.5cm]{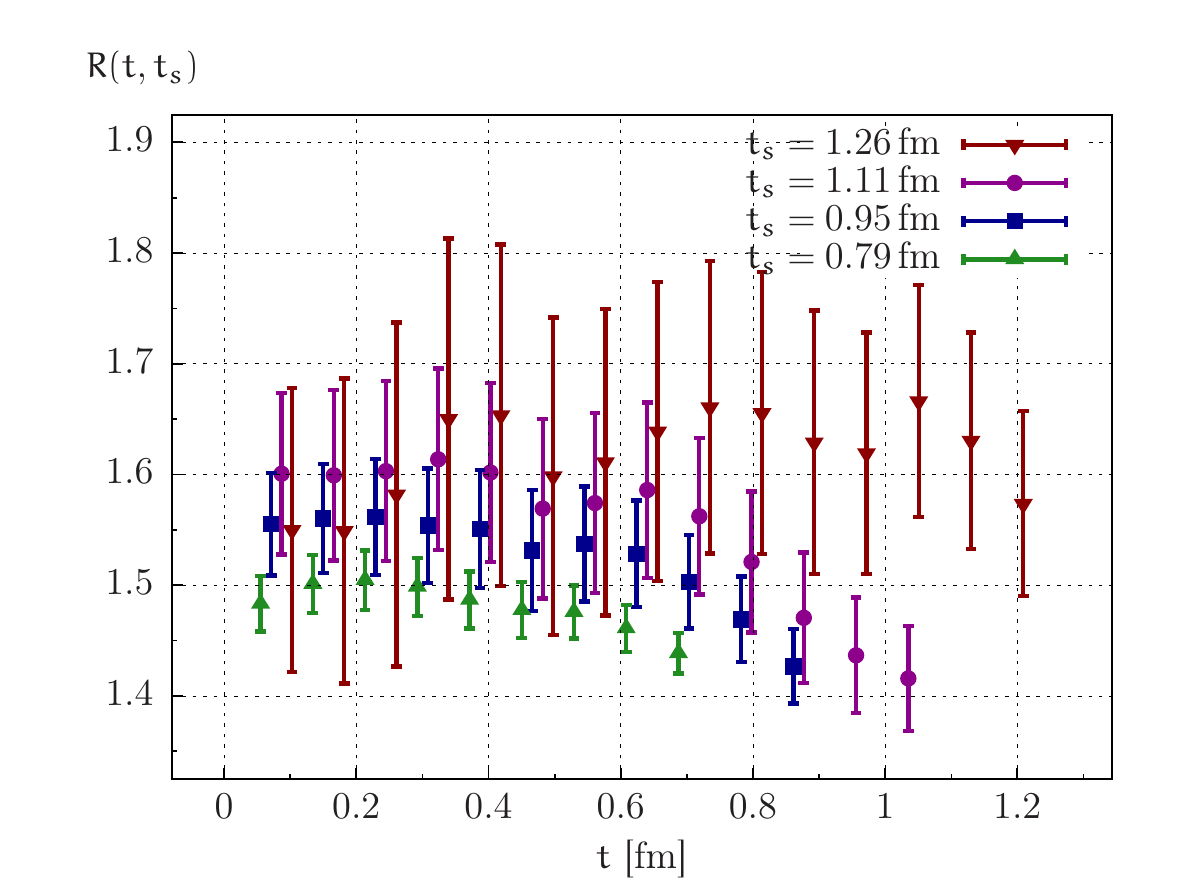}
\caption{\small The ratio $R(t,t_{\rm s})$ at $\beta=5.2$ and
  $m_\pi=312\,\MeV$ for several different values of the source-sink
  separation $t_{\rm s}$.}
\label{fig:A5ratio}
\end{center}
\end{figure}

The lowest-lying multi-particle state in the nucleon channel consists of one
nucleon and two pions forming an S-wave. Therefore, assuming that nucleon and
pions are only weakly interacting, one expects the gap $\Delta$ to be
proportional to the pion mass, since the mass difference to the Roper
resonance amounts to about 500 MeV. The same argument applies if one nucleon
and one pion form a P-wave, provided that the non-zero momentum induced by the
box size is small enough.  It is then clear that excited states may
increasingly distort the results for $\gA$ as the physical pion mass is
approached. A typical situation is shown in Fig.\,\ref{fig:A5ratio}: as
$t_{\rm s}$ is varied from $0.8\,\fm$ to $1.26\,\fm$, the ratio $R(t,t_{\rm
  s})$ is shifted by about 10\% to larger values. Given the rapid degradation
of the signal, it then remains unclear whether $t_{\rm s}\approx 1\,\fm$ is
sufficient to rule out a bias in the result for $\gA$. The most
straightforward strategy to address this problem is to include the first
excitation into the fit {\it ansatz} for $R(t,t_{\rm s})$ (see
ref.\,\cite{JGreen_lat11}) or to investigate larger values of $t$ and $t_{\rm
  s}$ \cite{ETMC_exc2011}.

\begin{figure}
\begin{center}
\includegraphics[width=8.5cm]{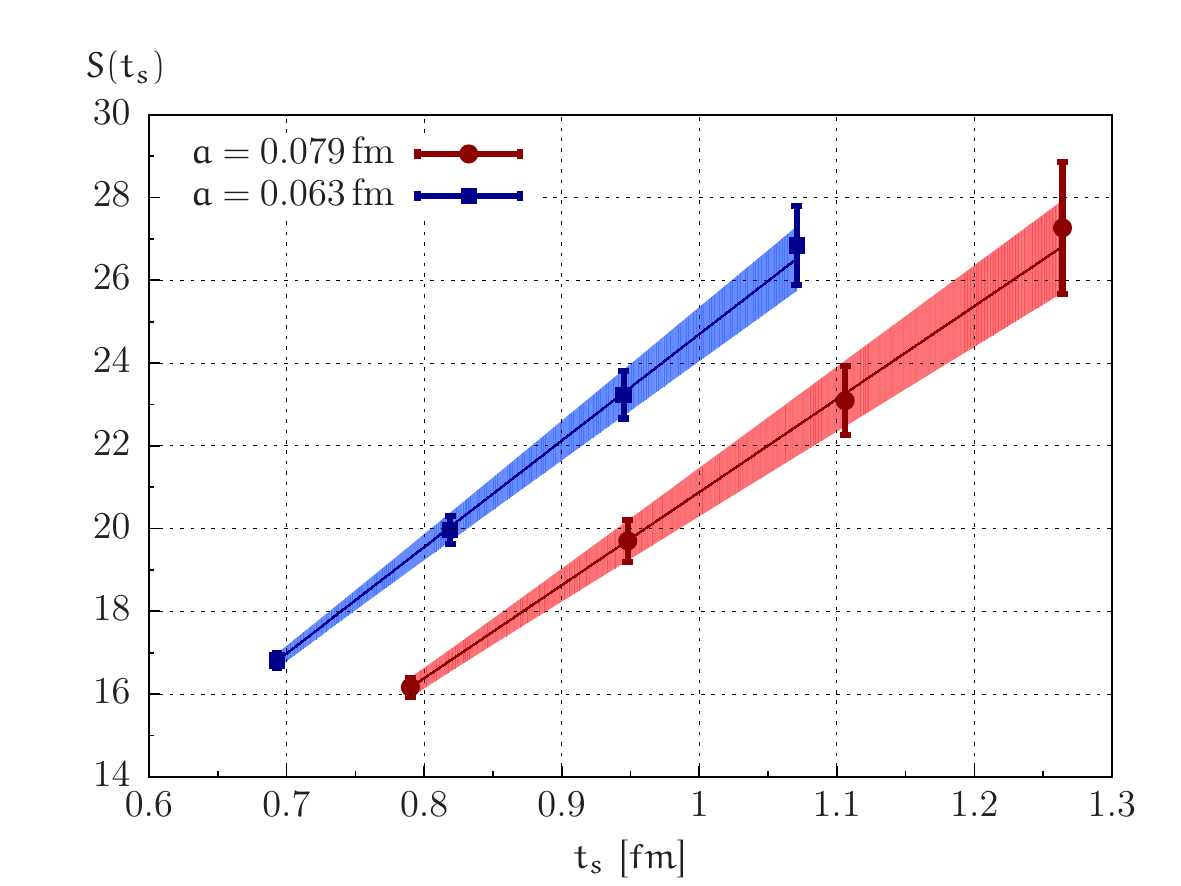}
\caption{\small The summed ratio $S(t_{\rm s})$ at $m_\pi\approx 320\,\MeV$
  for two different lattice spacings (ensembles A5 and F6).}
\label{fig:Sratio}
\end{center}
\end{figure}

Here we present an alternative approach, based on the use of summed
operator insertions
\cite{Maiani:1987by,smear:Gaussian89,Bulava:2011yz}. The key
observation is that excited state contributions can be parametrically
reduced when $R(t,t_{\rm s})$ is summed over~$t$. More precisely, the
asymptotic behaviour of the summed ratio $S(t_{\rm s})$ is given by
\begin{equation}
   S(t_{\rm s}) := \sum_{t=0}^{t_{\rm s}} R(t,t_{\rm s})\;
  \stackrel{t_{\rm s}\gg0}{\longrightarrow}\; c+t_{\rm s}
   \left\{\gA^{\rm{bare}} 
  +\rmO(\rme^{-{\Delta}t_{\rm s}})\right\},
\label{eq:Sratio}
\end{equation}
where the (divergent) constant, $c$, includes contributions from
contact terms. By computing $S(t_{\rm s})$ for several sufficiently
large values of $t_{\rm s}$, the quantity of interest can be extracted
from the slope of a linear fit. Since $t_{\rm s}> t, (t_{\rm s}-t)$ by
construction, excited state contributions to the slope of
$S(t_{\rm{s}})$ are more strongly suppressed relative to
$R(t,t_{\rm{s}})$. Compared to the standard method of computing the
latter at a single fixed value of $t_{\rm{s}}$, it is clear, however,
that the approach via summed insertions is computationally more
demanding. In Fig.\,\ref{fig:Sratio} we show typical fits to the
summed ratio $S(t_{\rm s})$ which demonstrate that the linear
behaviour is very well satisfied.

\section{Results}

We have determined the axial charge by fitting the
summed correlator $S(t_{\rm{s}})$ to a linear function for
$0.7\,\fm\lesssim t_{\rm{s}}\lesssim 1.3\,\fm$ and multiplying the
slope by the relevant renormalization factor of the axial current,
eq.\,(\ref{eq:za}). We have verified the stability of the method by
excluding the smallest value of $t_{\rm{s}}$ from the fit for each
ensemble. Typically, this leads to an increase in the value for $\gA$,
albeit with a $1.5-2$ times larger statistical error. 

\begin{figure}
\begin{center}
\includegraphics[width=8.5cm]{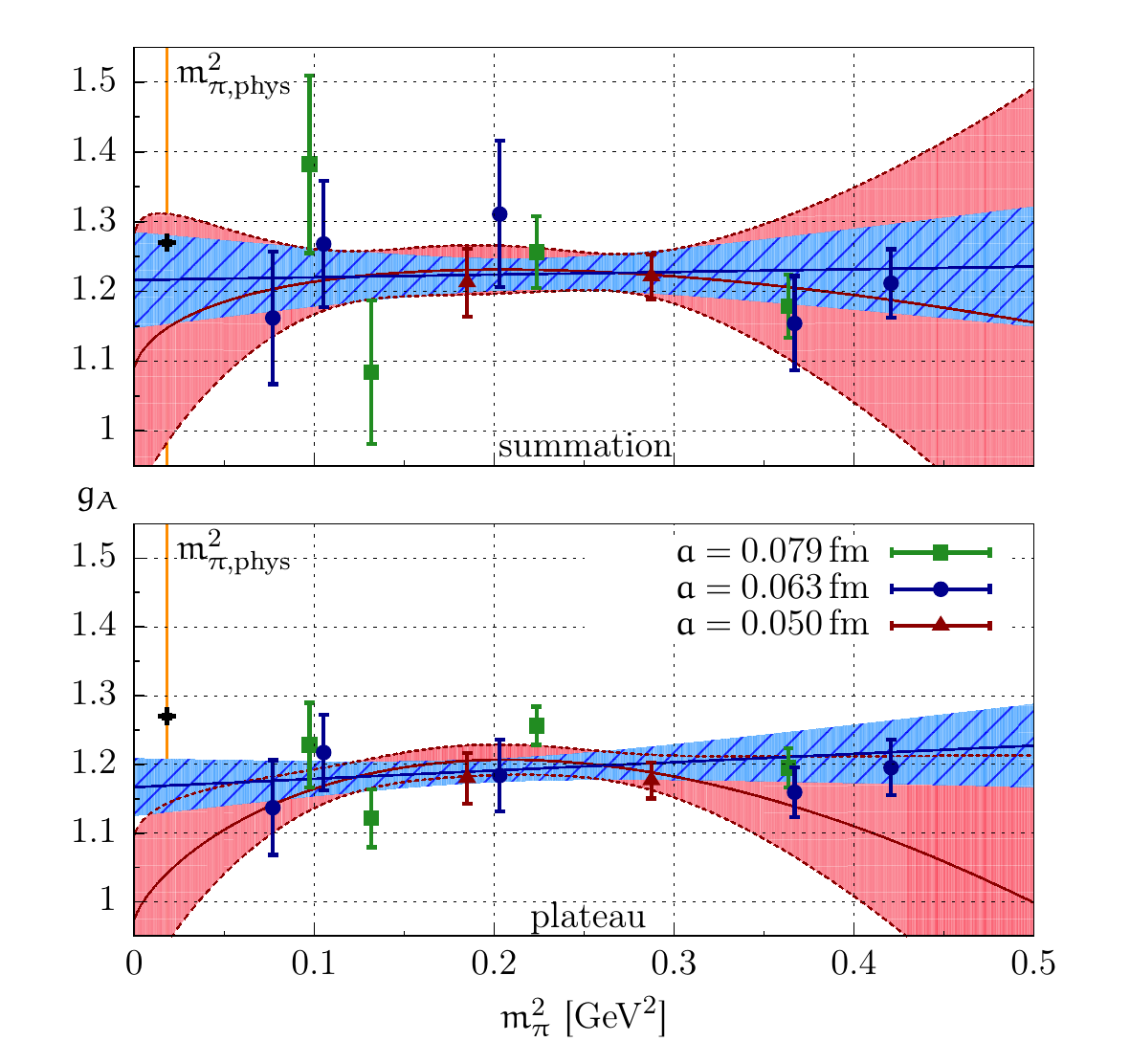}
\caption{\small Chiral behaviour of $\gA$ extracted from summed insertions
  (upper panel) and using the standard plateau method (lower
  panel). Chiral fits of type ``A'' and ``D'' (see text) applied for
  $m_\pi<540\,\MeV$ are represented by the blue/hatched and red bands,
  respectively. The black point denotes the experimental value.} 
\label{fig:gA}
\end{center}
\end{figure}

In the following we present a detailed comparison between the results
obtained using summed insertions (``summation method'') with those
arising from fitting the ratio $R(t,t_{\rm{s}})$ to a constant in~$t$
for $t_{\rm{s}}\approx 1.1\,\fm$ (``plateau method'').
Results are shown in Table\,\ref{tab_results} and Fig.\,\ref{fig:gA}. One
observes that the plateau method yields estimates for $\gA$ that mostly lie
below the experimental value, which is the typical behaviour seen in other
calculations at similar pion masses. Typically, the summation method produces
results which are higher than those from the plateau method, in some cases by
up to 10\%. At the same time, the summation method has larger statistical
errors. Still, since an increase is observed in seven out of eleven cases,
while a slightly smaller value was obtained only for one ensemble, it is
unlikely that this can be merely attributed to statistical fluctuations.

\begin{table}
\begin{ruledtabular}
\begin{tabular}{c c c c c c}
Run & $a [\fm]$ & $m_\pi [\MeV]$ & $m_{\rm N}/m_\pi$ &
 $\gA^{\rm summ}$ &  $\gA^{\rm plat}$ \\
\hline
A2 & 0.079 & 603 & 2.454(15) & 1.179( 45) & 1.195(28) \\
A3 &       & 473 & 2.803(23) & 1.256( 52) & 1.256(28) \\
A4 &       & 363 & 3.309(41) & 1.084(103) & 1.121(42) \\
A5 &       & 312 & 3.751(77) & 1.382(127) & 1.228(61) \\
\hline
E3 & 0.063 & 649 & 2.462(12) & 1.212( 49) & 1.195(40) \\
E4 &       & 606 & 2.561(11) & 1.154( 68) & 1.160(36) \\
E5 &       & 451 & 2.920(22) & 1.311(105) & 1.184(52) \\
F6 &       & 324 & 3.683(48) & 1.268( 91) & 1.217(55) \\
F7 &       & 277 & 3.771(86) & 1.162( 95) & 1.137(37) \\
\hline
N4 & 0.050 & 536 & 2.581(13) & 1.221( 32) & 1.176(26) \\
N5 &       & 430 & 2.939(28) & 1.212( 48) & 1.180(37) \\
\end{tabular}
\end{ruledtabular}
\caption{Results for the axial charge determined via summed insertions
  and the conventional plateau method. The data are corrected for
  finite-volume effects estimated in HBChPT (see text).}
\label{tab_results}
\end{table}

In order to investigate the chiral behaviour in detail, we have
performed chiral extrapolations based on several different {\it
ans\"atze} commonly used in the
literature\,\cite{nuclFF:QCDSF06_nf2,nuclFF:RBC08_nf2p1,nuclFF:LHPC10_nf2p1,
Colangelo:2010ba, nuclff:ETMC10_nf2}, i.e.
\begin{eqnarray}
  \hbox{Fit~A:} &\quad & \alpha +\beta m_\pi^2  \nonumber \\
  \hbox{Fit~B:} &\quad & \alpha^\prime +\beta^\prime m_\pi^2
  -|\gamma^\prime|\, m_\pi^2\ln m_\pi^2/\Lambda^2  \\
  \hbox{Fit~C:} &\quad & \alpha^{\prime\prime} +\beta^{\prime\prime}
  m_\pi^2 -|\gamma^{\prime\prime}|\, \rme^{-m_\pi L}, \nonumber
\end{eqnarray}
with fit parameters $\alpha, \beta, \alpha^\prime,\ldots$. Another {\it
  ansatz}, Fit~D, is a three-parameter fit, based on the expressions derived
in Heavy-Baryon Chiral Perturbation Theory (HBChPT) in infinite
volume\,\cite{Hemmert:2003cb,Beane:2004rf}, with three additional low-energy
constants fixed by phenomenology\,\cite{nuclff:ETMC10_nf2}. Examples are shown
in Fig.\,\ref{fig:gA}. A simple linear chiral extrapolation (Fit~A) applied to
the data from all three lattice spacings for which $m_\pi<540\,\MeV$ yields a
value for $\gA$ at the physical pion mass which agrees well with experiment
within the statistical uncertainty. A similar statement applies to the fit
based on HBChPT (Fit~D). By contrast, extrapolations of the data determined
using the plateau method fail to reproduce the experimental value by two
standard deviations.

\begin{figure}
\begin{center}
\includegraphics[width=8.5cm]{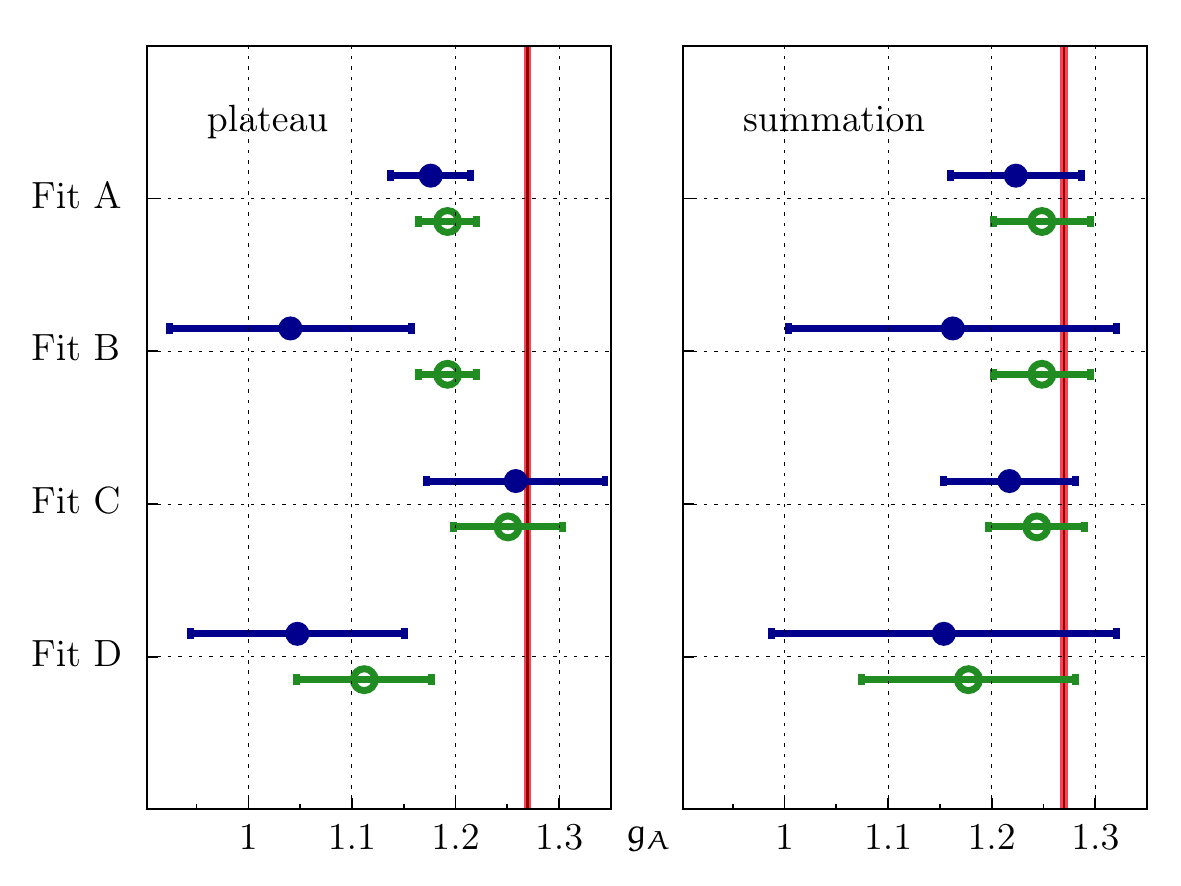}
\caption{\small Results for $\gA$ at the physical pion mass for the
  plateau and summation methods. Solid points refer to a pion mass cut
  at $m_\pi<540\,\MeV$, while open symbols are used to denote results
  from fits across the entire pion mass range. Fits A, B and~D were
  applied to the volume-corrected data (see text). The vertical lines
  represent the experimental value.}
\label{fig:gAcomp}
\end{center}
\end{figure}

Fit~C was introduced in \cite{nuclFF:RBC08_nf2p1} to test whether the
widely observed underestimates of $\gA$ could be a manifestation of
finite-volume effects. After determining the parameters
$\alpha^{\prime\prime}, \beta^{\prime\prime}$ and
$\gamma^{\prime\prime}$, the volume-dependent term proportional to
$\exp\{-m_\pi L\}$ can be subtracted. Indeed, a non-zero value for
$\gamma^{\prime\prime}$ results when fit~C is applied to the data
obtained via the plateau method. A linear chiral extrapolation, using
the fitted coefficients $\alpha^{\prime\prime}$ and
$\beta^{\prime\prime}$, then yields an estimate for $\gA$ which agrees
with experiment (see Fig.\,\ref{fig:gAcomp}). However, repeating the
procedure for the summation method produces a vanishing coefficient
$\gamma^{\prime\prime}$. We conclude that, in this case, there is no
need to subtract any term designed to account for finite-volume
effects, in order to get agreement with experiment. When addressing
the influence of excited states it is important to realize that such
contributions are volume-dependent, whenever they are due to
multiple-particle states. Thus, for a true benchmark calculation of
the axial charge one must be able to separate finite-volume
corrections to $\gA$ from volume-dependent excited-state
contamination.

Figure\,\ref{fig:gAcomp} shows a compilation of the chirally
extrapolated $\gA$ from the four different fit types. While summed
insertions invariably produce estimates that are compatible with
experiment, one consistently obtains lower values using the plateau
method, except for fit~C with the term containing $\exp\{-m_\pi L\}$
subtracted. These observations are stable under variations of the pion
mass range, as indicated in the figure.

We now proceed to discussing our final result and the estimation of
systematic errors. We applied a finite-volume correction based on the
expression derived in HBChPT \cite{Hemmert:2003cb} (see
ref.\,\cite{nuclff:ETMC10_nf2} for details on the numerical
evaluation).  Since $m_\pi L>4$ and $2\,\fm\leq L\leq 3\,\fm$ the
resulting shifts are at the sub-percent level for all our
ensembles. As our best estimate, we quote the result from fit~A,
applied to the volume-corrected data obtained from summed insertions,
with a cut of $m_\pi < 540\,\MeV$, i.e.
\begin{equation}
   \gA=1.223\pm0.063\,(\hbox{stat}),
\label{eq:gAstat}
\end{equation}
which agrees with the PDG average\,\cite{PDG10} of $1.2701(25)$ within
the statistical error. By contrast, when the same fitting procedure is
applied to the results extracted from the plateau method, one finds
the much lower estimate of $\gA=1.173\pm0.038\,(\rm stat)$.

\begin{figure}
\begin{center}
\includegraphics[width=8.5cm]{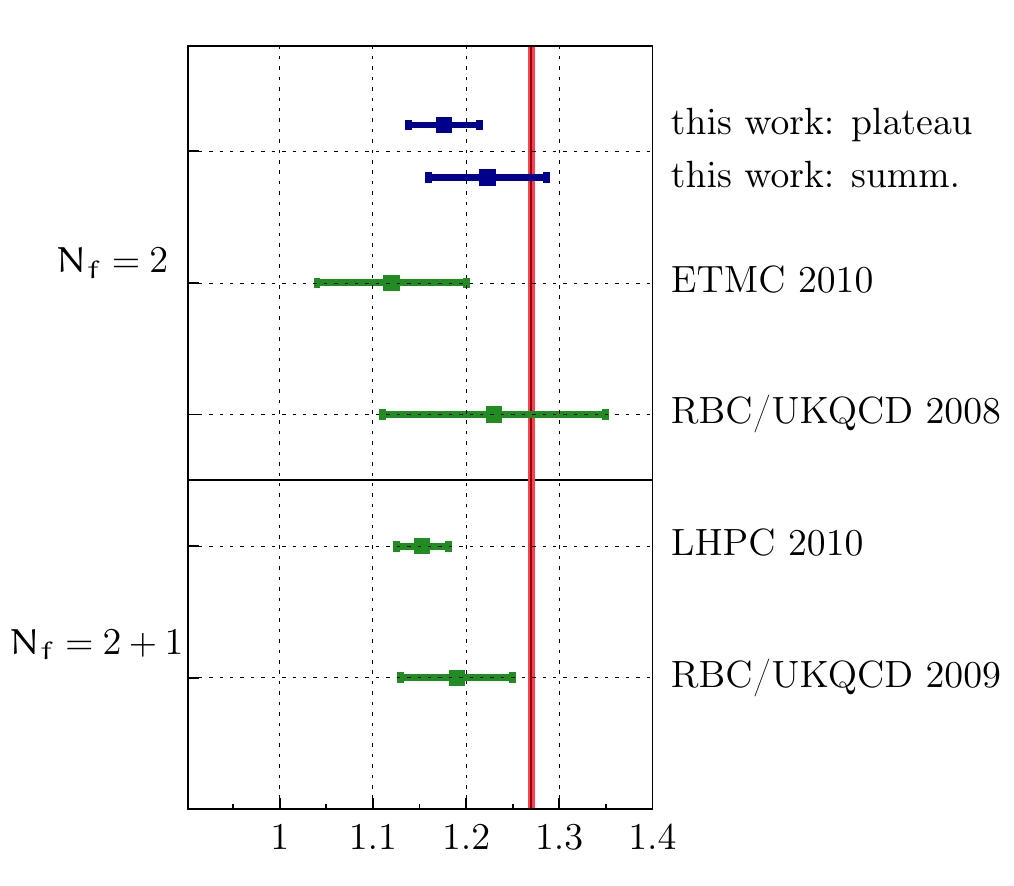}
\caption{\small Estimates for $\gA$ determined from the summation and plateacu
  methods compared to recent results by ETMC \cite{nuclff:ETMC10_nf2},
  RBC/UKQCD (\cite{nuclFF:RBC08_nf2} for $\Nf=2$,
  ref. \cite{nuclFF:RBC09_nf2p1} for $\Nf=2+1$) and LHPC
  \cite{nuclFF:LHPC10_nf2p1}. Only statistical errors are shown. The thick
  vertical line represents the experimental result.}
\label{fig:compilation}
\end{center}
\end{figure}

It is instructive to compare our findings to other recent results for the
axial charge. A compilation is plotted in Fig.\,\ref{fig:compilation}. With
the exception of the results by
RBC/UKQCD\,\cite{nuclFF:RBC08_nf2,nuclFF:RBC09_nf2p1}, our estimate based on
the summation method is the only one which agrees with the experimental value
within statistical errors. It it also worth mentioning that, in order to
achieve agreement with experiment, a large downward curvature in the data had to be
separated off in refs.\,\cite{nuclFF:RBC08_nf2,nuclFF:RBC09_nf2p1}, by
applying the procedure of fit~C.

We note that, with our current level of statistical accuracy, no
significant dependence on the lattice spacing could be detected. For
instance, applying fits A--D only to the data at $\beta=5.3$ produces
a tiny variation, which is 10 times smaller than the statistical
error. Therefore we refrain from quoting a separate systematic
uncertainty relating to cutoff effects.  In order to quantify the
uncertainty associated with the chiral extrapolation, we adopted two
procedures. First, by applying different cuts to the upper limit on
the pion mass interval between~470 and 640\,\MeV, we observe a
variation of $\pm0.035$ relative to the central value in
eq.\,(\ref{eq:gAstat}). Second, we considered the spread among fits
A--D as a measure for the uncertainty relating to the extrapolation,
which amounts to a downward shift by $-0.060$. Taking the largest
upward and downward variations from both methods as the error
estimate, we arrive at our final result
\begin{equation}
   \gA=1.223\pm0.063\,(\hbox{stat})\,
   {}^{+0.035}_{-0.060}\,(\hbox{syst}), 
\label{eq:gAfinal}
\end{equation}
which agrees with the experimental result at the level of $6-7\%$.

\section{Conclusions}

The typical source-sink separations in baryonic three-point functions can be
smaller by up to a factor two compared to those used in the mesonic
sector. Even for $t_{\rm s}\approx1.3\,\fm$ it is hard to judge whether or not
a significant bias due to excited state contributions can be excluded, if the
standard plateau method is employed without further checks (see
Fig.\,\ref{fig:A5ratio}). Summed operator insertions offer an attractive
alternative, since excited state contributions are parametrically more
strongly suppressed relative to those encountered in conventional ratios. Our
findings, summarized in Fig.\,\ref{fig:gAcomp}, demonstrate that a much better
agreement with the experimental value of $\gA$ can be achieved in this way. On
the downside, one must list the necessity to compute correlation functions for
several source-sink separations, as well as the larger statistical errors
associated with the method. However, since excited state contamination might
be a generic problem for lattice calculations of structural properties of the
nucleon, the larger numerical effort seems a worthwhile investment. We plan to
corroborate our findings by including additional ensembles with smaller pion
masses and extend our studies to other quantities, such as the vector and
axial vector form factors of the nucleon. For this purpose, optimised
anisotropic smearing functions for non-vanishing hadron
momenta\,\cite{Roberts:2012tp,DellaMorte:2012xc} may prove to be a useful
addition to the technique of summed insertions.

\begin{acknowledgments}
Our calculations were performed on the ``Wilson'' HPC Cluster at the Institute
for Nuclear Physics, University of Mainz. We are grateful to Dalibor
Djukanovic for technical support. This work was supported by DFG (SFB\,443 and
SFB\,1044) and the Rhineland-Palatinate Research Initiative. We are grateful
to our colleagues within the CLS initiative for sharing ensembles.
\end{acknowledgments}

\end{document}